\documentstyle[aps,epsf,preprint]{revtex}
\tightenlines
\title{Pseudogaps: A third peak in the fermion spectral function.}

\author{Oleg Tchernyshyov}

\address{Physics Department, Columbia University, New York, New York 10027}

\preprint{
To appear in Phys.~Rev.~B}
\begin{document}

\maketitle

\begin{abstract}  
I present an exactly solvable model of a pseudogap with two 
zero-energy fermion modes 
coupled to each other by a classical source of frequency $\omega_0$ and 
strength $|\Delta|$.  A suitably defined fermion propagator has 
an infinite number of poles at frequencies that are multiple integers of
$\omega_0$.  In the adiabatic limit, $\omega_0\ll|\Delta|$, 
the situation is {\em qualitatively} different from the static
case $\omega_0=0$: the residue of the pole at $\omega=0$ (a remnant of the 
bare fermion) vanishes linearly with $\omega_0$, a result that could not
be anticipated by perturbation theory; the multiple poles of the propagator
coalesce into a continuum instead of forming two single poles at 
$\pm|\Delta|$, which should be interpreted as inhomogeneous broadening
of the Bogoliubov quasiparticles.  
\end{abstract}


The notion of a pseudogap is relevant to physics of fermionic systems 
with slowly fluctuating excitations of the Bose type.  In a state with 
long-range order, fermions are strongly scattered by a condensate of
these excitations, which splits the Landau quasiparticle peak into two
and thus creates a gap in the fermion energy spectrum.  In the disordered
state, slow fluctuations can produce a qualitatively similar effect.  
Underdoped cuprate superconductors offer an example of such behavior.   
Without carrier doping, the cuprates are antiferromagnetic (AFM) insulators.  
Although hole doping destroys the AFM long-range order, slow spin
fluctuations create a remnant of the Mott-Hubbard energy gap (a few tenths 
of an electronvolt \cite{ARPES}).  There is very little spectral weight at 
lower energies, which makes these materials rather poor conductors.    

By using a low-level approximation to estimate the fermion self-energy 
(a bare intermediate fermion and a reasonably guessed particle-particle
susceptibility), Kampf and Schrieffer
\cite{Kampf} have found an interesting feature 
in the fermion spectral function in a pseudogap regime.  
In addition to two peaks at energies 
$\pm \tilde{\epsilon}_{\bf k}\equiv \pm\sqrt{\epsilon_{\bf k}^2+|\Delta|^2}$ 
(as expected in the ordered state), they found a small {\em third} peak at 
the old location $\epsilon_{\bf k}$, a remnant of the Fermi-liquid state.  As 
the fluctuations slow down, the third peak gradually disappears.  If true,
this finding could give an insight into the formation of 
{\em coherent} spectral features (i.e., fermionic quasiparticles) in 
the cuprates upon doping.    

Deisz, Hess, and Serene \cite{Serene} have attempted to verify these results
in the conserving FLEX approximation \cite{FLEX} and found {\em no} 
midgap peak.  In our opinion, this negative result is related to 
a particular form of the self-consistently calculated spin susceptibility:
while in \cite{Kampf} it was assumed that magnons have a typical {\em real} 
frequency $\omega_0$, the self-consistently calculated susceptibility
\cite{Serene} clearly has a diffusing form.  Whether or not propagating 
magnons can exist in the pseudogap regime is certainly an open question.  
Even if they can, it is far from clear whether the midgap band is a robust
feature or an artifact of a low-level approximation.  

With this in mind, I have studied a toy model which contains the
essential features of the Kampf--Schrieffer theory, yet permits an exact 
solution.  Basing on the results obtained within this model, one can make
the following conclusions.  

(a) The approximation of Kampf and Schrieffer gives reliable 
results in the limit of {\em fast} fluctuations $\omega_0/|\Delta|\gg 1$,
which corresponds to a Fermi-liquid state.   

(b) The quasiparticle peak at $\omega=0$ persists well into
the pseudogap state ($\omega_0/|\Delta|\ll 1$).  Its spectral weight vanishes 
as $\omega_0/|\Delta|$, which is a non-perturbative result.  

(c) In the adiabatic limit $\omega_0\to 0$, the two 
quasiparticle peaks at $\pm \tilde{\epsilon}_{\bf k}$ are {\em strongly}
broadened.  Broken ergodicity is necessary to restore their sharpness.  

(d) When the energies of the coupled fermion modes are not equal, the 
spectral function in the adiabatic limit is a continuum with three peaks,
none of which is a pole of the fermion propagator.  

The toy model is defined as follows.  Two fermion modes $a_1$ and $a_2$
interact with each other via a classical, time-dependent external source
whose frequency equals $\omega_0$; the Hamiltonian of this system is 
\begin{equation}
H = (\Delta a_1^\dagger a_2 +\Delta^* a_2^\dagger a_1)\cos{\omega_0 t}.
\label{ham}
\end{equation}
By an appropriate unitary transformation, $\Delta$ can always be made
positive, $\Delta=\Delta^*=|\Delta|$.  It should be noted that a similar
model of two fermion modes interacting with a single (quantum) boson mode
of frequency $\omega_0$ has been considered by Schrieffer \cite{Schrieffer}.  

If the external source is static, $\omega_0=0$, the Hamiltonian is readily 
diagonalized: there are two eigenmodes $a_\pm\equiv (a_1\pm a_2)/\sqrt{2}$
with energies $\mp|\Delta|$.  
When $\omega_0\neq 0$, symmetry with respect to time translations is 
broken and the time-ordered propagator matrix
$G_{\alpha\beta}(t,t') 
= -i\langle\psi|T[a_\alpha(t) a^\dagger_\beta(t')]|\psi\rangle$
becomes a function of two time variables.  In order to restore this 
symmetry (and conservation of energy with it), I perform translations 
$t\to t+\tau$, $t'\to t'+\tau$
and average $G(t+\tau,t'+\tau)$ with respect to $\tau$.  This process
is similar to restoring translational invariance in the impurity problem
by averaging over impurity positions throughout the crystal.
The resulting propagator 
\begin{equation}
\overline{G}(t-t') = \langle G(t+\tau,t'+\tau)\rangle_\tau
\label{G bar defined}
\end{equation}
is a function of $t-t'$ only.  
Treating $H$ as a perturbation, one can write out a series 
for $G(t,t')$ by iterating the matrix Dyson equation 
\[G(t,t') = G^{(0)}(t-t') 
+ |\Delta|\int dt''\ G^{(0)}(t-t'')\ \sigma_1\cos{\omega_0t''}\ G(t'',t').\]
Upon averaging over time shifts, diagrams with an odd number of
cosine factors vanish, making the series for $\overline{G}$ diagonal
(this would not be so in the static case):
\begin{eqnarray}
\overline{G}(t-t') &=& G^{(0)}(t-t') \nonumber\\
&+& \frac{|\Delta|^2}{2}\int dt_1 dt_2\ 
G^{(0)}(t-t_2)\ G^{(0)}(t_2-t_1)\cos{\omega_0(t_2-t_1)}\ G^{(0)}(t_1-t') 
\label{G bar}\\
&+&\ldots\nonumber
\end{eqnarray}
A generic term of the perturbation series for $\overline{G}$ can be readily 
written out, most easily for the Fourier transform of (\ref{G bar}) 
$\overline{G}(\omega)$ --- see Fig.~\ref{graphs}.
Note that the effective coupling constant is $g\equiv|\Delta|^2/4$.  

To second order in $|\Delta|$, the self-energy is
\[
\Sigma^{(2)}(\omega)
= \frac{|\Delta|^2}{4(\omega-\omega_0)}
+ \frac{|\Delta|^2}{4(\omega+\omega_0)} 
= \frac{|\Delta|^2\omega}{2(\omega^2-\omega_0^2)}. 
\]
It is similar to $\Sigma$ of Kampf and Schrieffer in that it
vanishes with a negative slope at $\omega=0$.  
Therefore, to this order,  $\overline{G}(\omega)=1/[\omega-\Sigma(\omega)]$ 
has {\em three} poles, at $\omega=\pm\sqrt{\omega_0^2+|\Delta|^2/2}$ and at
$\omega=0$, the latter with the residue 
\begin{equation}
z_0 = \frac{\omega_0^2}{\omega_0^2+|\Delta|^2/2},
\label{z_0}
\end{equation}
which apparently vanishes as $2\omega_0^2/|\Delta|^2$ in the limit of
slow fluctuations.  Simultaneously, the other two poles are approaching
$\pm|\Delta|/\sqrt{2}$ and their residues tend to $1/2$, as in the static 
case.  When fluctuations are fast, the dominant pole is at $\omega=0$.  Thus
a picture of a smooth crossover from one to two peaks in the spectral 
function emerges.  

This smoothness is, however, misleading.  As has already been 
noted, there are off-diagonal contributions to the propagator matrix
in the static case, but not in the fluctuating case.  Therefore,
the adiabatic limit ($\omega_0\to 0$) is {\em not} 
expected to resemble the static situation ($\omega_0=0$).  
In addition, higher powers of $|\Delta|$
in the self-energy can be neglected only when $|\Delta|$ is small, which
for low fermion frequencies $\omega$ translates into 
$|\Delta|\ll\omega_0$.  Another indication of a nonperturbative character 
of the adiabatic limit comes from the exact expression for the residue 
of the $\omega=0$ pole obtained below.  As $g\equiv|\Delta|^2/4\to \infty$,
the behavior of $z_0$ is nonanalytic: $z_0\sim {\rm const\ }g^{-1/2}$.

{\bf Exact solution.}  By making a unitary transformation, the Hamiltonian 
(\ref{ham}) can be diagonalized (in the same way as in the $\omega_0=0$ case):
\[H = |\Delta|\cos{\omega_0 t}\ (a_-^\dagger a_- - a_+^\dagger a_+). \]
Time dependence of the annihilation operators is given by 
\[a_\pm(t) = a_\pm(0)
\exp{\left(\pm i\frac{|\Delta|}{\omega_0}\sin{\omega_0t}\right)}.\]
After choosing a state vector $|\psi\rangle$, one can immediately write down 
the propagator $\overline{G}(t-t')$.  A particular choice of $|\psi\rangle$ 
does not affect the form of $\overline{G}(\omega)$ but rather determines the 
integration path in the complex plane of $\omega$.  If we select 
the state annihilated by $a_-$ and $a^\dagger_+$, 
\[
G_{\pm}(t,t') = \mp i
\exp{\left[\pm i\frac{|\Delta|}{\omega_0}
(\sin{\omega_0t}-\sin{\omega_0t'})\right]}
\theta(\pm t \mp t')
\]
and, after averaging over time shifts, 
\[
\overline{G}_{\pm}(t) = \mp i
J_0\left(\frac{2|\Delta|}{\omega_0}
\sin{\frac{\omega_0 t}{2}}\right)\theta(\pm t),
\]
where $J_0(x)$ is a Bessel function.  Note that $G(t,t')$ and 
$\overline{G}(t-t')$ behave differently in the limit $\omega_0\to 0$.  
Without averaging, fermions $a_\pm$ have well-defined energies in this
limit.  By averaging over time, we force them to slowly sample the 
whole range of energies between $-|\Delta|$ and $|\Delta|$.  Evidently, 
this would take an infinite amount of time as $\omega_0\to0$.  

The Fourier transform $\overline{G}(\omega)$ can now be readily determined.  
In the limit $\omega_0\to0$, 
$\overline{G}(\omega) \to 1/\sqrt{\omega^2-|\Delta|^2}$
with all of the spectral weight 
\begin{equation}
\overline{{\cal A}}(\omega) \to \overline{{\cal A}}_0(\omega)
\equiv \frac{1}{\pi\sqrt{|\Delta|^2-\omega^2}}
\label{static A bar}
\end{equation}
concentrated between 
$-|\Delta|$ and $|\Delta|$.  This result can be simply understood. 
On a time scale less than $1/\omega_0$, each individual fermion 
has a well-defined frequency $\omega$ equal to the instanteneous
amplitude of the external source.  Since the 
latter fluctuates as $|\Delta|\cos{\omega_0 t}$,
the probability density to find a certain strength $\omega$ of this field 
is given precisely by (\ref{static A bar}).  Thus the density of fermions
whose pole is shifted from 0 to $\omega$ is (\ref{static A bar}).  
Similar broadening in the limit of slow fluctuations has been discussed
in connection with a pairing \cite{Schmid} and Peierls \cite{Sadovskii} 
pseudogaps.  This simple observation could perhaps vindicate the FLEX
approximation in light of its recent criticism \cite{Vilk}: the strong 
broadening of quasiparticle peaks by slow AFM fluctuations is a natural 
phenomenon and not an artifact of the FLEX approximation.  

\begin{figure}
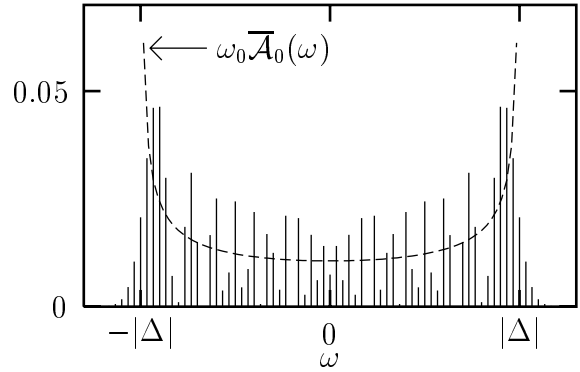

\caption{Density of states for $|\Delta|/\omega_0=30$.  Height of impulses 
represents spectral weight of discrete levels.  The broken 
line is the result of an adiabatic extrapolation.}
\label{dos}
\end{figure}

When $\omega_0\neq 0$, the spectral weight of an $a_\pm$ fermion is a 
superposition of sharp peaks 
$\overline{{\cal A}}(\omega) = \sum_n z_n\delta(\omega - n\omega_0)$ 
with residues $z_n = J_n^2(|\Delta|/\omega_0)$.  (This result differs 
markedly from a Poisson distribution found in the case of a quantum 
boson \cite{Schrieffer}.)  The spectral weight 
at $\omega=0$ is a remnant of the bare fermion.  In the fast limit,
$z_0\sim 1-|\Delta|^2/2\omega^2$, which agrees with the perturbative 
expansion for $z_0$ (\ref{z_0}).  In the slow limit, 
\begin{equation}
z_0 \approx \frac{\omega_0[1+\sin{(2|\Delta|/\omega_0)}]}{\pi|\Delta|},
\label{asymptotic z_0}
\end{equation}
which certainly does {\em not} agree with (\ref{z_0}).  Apart from the
oscillatory part, $z_0$ in this limit can be obtained as the area
under the adiabatic density of states (\ref{static A bar}) between
$-\omega_0/2$ and $\omega_0/2$, as illustrated in Fig.~\ref{dos} 
(note that chaotic oscillations of the peak heights are absent 
close to $\omega=\pm|\Delta|$).  Thus, unlike the perturbation theory, 
the picture of adiabatic evolution provides a reasonable description of 
the system in the limit of slow fluctuations.  

\begin{figure}
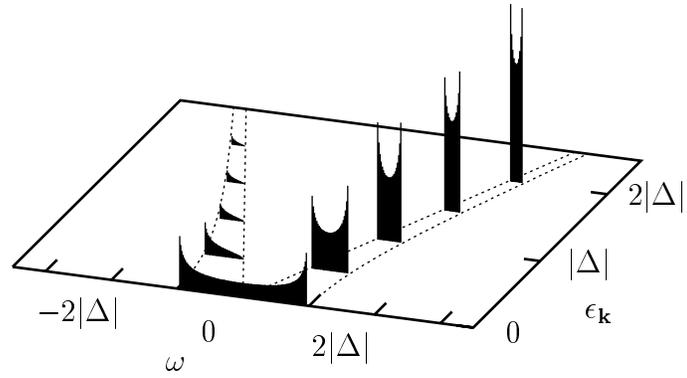

\caption{Fermion spectral weight $\overline{{\cal A}}(\omega)$ in the 
adiabatic limit for positive bare energies $\epsilon=0\ldots 2|\Delta|$.
Dotted lines are $\omega=\pm\epsilon$ and $\omega=\pm\tilde{\epsilon}$.}
\label{weight}
\end{figure}

{\bf Fermions with nonzero bare energy: adiabatic limit.}  When the bare 
fermion modes $a_{1,2}$ have unequal energies $\epsilon_{1,2}=\pm\epsilon$, 
the problem does not
admit such a simple solution for a finite $\omega_0$ because the 
Hamiltonian does not commute with itself at different times.  However, it
is still possible to find the $\tau$-averaged propagator
(\ref{G bar defined}) in the adiabatic limit $\omega_0\to0$.  To do so, we 
first determine the propagator for a given instanteneous 
strength of the external source $\delta=|\Delta|\cos{\omega_0 t}$,
\[G(\omega,\delta) = \frac{1}{\omega^2-\epsilon^2-\delta^2}
\left(\begin{array}{cc}
\omega-\epsilon & \delta \\
\delta & \omega+\epsilon
\end{array}\right),\] 
and then average it over $\delta$ using the probability density 
(\ref{static A bar}).  As a result, off-diagonal (anomalous) components
$\overline{G}_{12}$ and $\overline{G}_{21}$ vanish, while 
\[\overline{G}_{11}(\omega) 
= \int_{-|\Delta|}^{|\Delta|} \frac{d\delta}{\pi\sqrt{|\Delta|^2-\delta^2}}\ 
\frac{\omega+\epsilon}{\omega^2-\epsilon^2-\delta^2}= 
\sqrt{\frac{\omega+\epsilon}{(\omega-\epsilon)(\omega^2-\tilde{\epsilon}^2)}}.
\]
with a nonzero spectral weight in the two ranges 
$\epsilon^2<\omega^2<\tilde{\epsilon}^2\equiv\epsilon^2+|\Delta|^2$:
\begin{equation}
\overline{{\cal A}}_{11}(\omega) = \frac{1}{\pi}
\sqrt{\frac{\omega+\epsilon}{(\omega-\epsilon)(\tilde{\epsilon}^2-\omega^2)}}.
\label{A11}
\end{equation}
For $\epsilon\neq 0$, there are {\em three} peaks in the spectral function, at
$\omega=\epsilon$ and $\pm\tilde{\epsilon}$.  None of these is a pole of
the propagator, but rather an inverse square-root branch point.  In the limit 
$\epsilon\gg|\Delta|$, the cut near $\epsilon$ evolves into a simple pole,
while the other cut vanishes (Fig.~\ref{weight}).  

At zero temperature, the occupation number of a fermion mode with bare energy 
$\epsilon$ is 
\[n(\epsilon) = \int_{-\infty}^{0}\overline{{\cal A}}_{11}(\omega)d\omega
= 1/2 - (\epsilon/\pi\tilde{\epsilon})\ 
{\bf K}'(\epsilon/\tilde{\epsilon}),\]
where ${\bf K}'(k)$ is a complete elliptic integral of the first kind. 
There is no discontinuity as $\epsilon$ crosses zero, although the slope
of $n(\epsilon)$ diverges logarithmically as $\epsilon\to0$.  
Assuming a constant bare density of fermion states (DOS), it
is also possible to obtain the DOS in the adiabatic limit by 
integrating (\ref{A11}) over $\epsilon$.  The resulting DOS
is identical to that of a pure $d$-wave superconductor, 
it vanishes linearly with $\omega$ in the middle of 
the pseudogap and diverges logarithmically at its edges.  

The simple model considered in this note indicates that the midgap band
of Kampf and Shrieffer may in fact be inseparable from the two bands above 
and below the pseudogap.  This effect can be ascribed to the fluctuations of
the gap amplitude in this model.  Calculations of the FLEX type
inevitably contain such fluctuations (see, e.g., \cite{Schmid})
and therefore do not lead to a sharp energy gap.  In principle, this 
adiabatic broadening of the quasiparticle peaks can be lifted if the 
observation time is shorter than the inverse frequency of the fluctuations 
$1/\omega_0$ (but still longer than $1/|\Delta|$).  If
fluctuations can also propagate in space, one has to limit
observations to a spatial area without significant retardation.  Given the
speed of the propagating fluctuations $s$, the size
of such an area should not exceed $s/\omega_0$.   

I acknowledge financial support from NEDO (Japan) 
and the US Department of Energy.



\begin{figure}
\caption{Feynman diagrams for $\overline{G}(\omega)$ through 
second order in the coupling constant $g\equiv|\Delta|^2/4$.
Broken lines: $|\Delta|/2$.  Solid lines: bare propagators $G^{(0)}$ 
at a frequency shown on the left.}
\label{graphs}
\end{figure}

\vskip 1cm

\vskip 1cm

\end{document}